# A Thermal-Photovoltaic Device Based on Thermally Enhanced Photoluminescence

Assaf Manor[1] and Carmel Rotschild[1,2]

[1] *Russell Berrie Nanotechnology Institute, Technion − Israel Institute of Technology, Haifa 32000, Israel*

[2] *Department of Mechanical Engineering, Technion − Israel Institute of Technology, Haifa 32000, Israel*

**Abstract**

Single-junction photovoltaic cells are considered to be efficient solar energy converters, but even ideal cells cannot exceed the their fundamental thermodynamic efficiency limit, first analysed by Shockley and Queisser (SQ)[1]. For moderated irradiation levels, the efficiency limit ranges between 30%-40%. The efficiency loss is, to a great extent, due to the inherent heat-dissipation accompanying the process of electro-chemical potential generation. Concepts such as solar thermo-photovoltaics[2,3] (STPV) and thermo-photonics[4] aim to harness this dissipated heat, yet exceeding the SQ limit has not been achieved, mainly due to the very high operating temperatures needed. Recently[5], we demonstrated that in high-temperature endothermic-photoluminescence (PL), the photon rate is conserved with temperature increase, while each photon is blue shifted. We also demonstrated how endothermic-PL generates orders of magnitude more energetic-photons than thermal emission at similar temperatures. These new findings show that endothermic-PL is an ideal optical heat-pump. Here, we propose and thermodynamically analyse a novel device based on Thermally Enhanced Photo Luminescence (TEPL). In such a device, solar radiation is harvested by a low-bandgap PL material. In addition to the PL excitation, the otherwise lost heat raises the temperature and allows the TEPL emission to be coupled to a higher bandgap solar cell. The excessive thermal energy is then converted to electrical work at high voltage and enhanced efficiency. Our results show that such a TEPL based device can reach theoretical maximal efficiencies of 70%, as high as in STPV, while the significantly lowered operating temperatures are below 1000C. In addition to the theoretical analysis, we experimentally demonstrated enhanced photo-current in TEPL device pumped by sub-bandgap radiation. This opens the way for a new direction in photovoltaics.

The ability of photoluminescence (PL) to extract thermal energy from a hot body was recently theoretically and experimentally demonstrated[5]. The Thermally-Enhanced PL (TEPL) radiation is characterized by a non-zero chemical potential at elevated temperatures, causing it to emit orders of magnitude more photons per second than the equal-temperature thermal body. This ability to exceed the black-body emission limit could be used for efficient conversion of solar-energy to electricity. In existing concepts such as Solar-Thermal Photovoltaics (STPV)[2,6] the solar heat flux is first converted to thermal emission by a selective absorber/emitter, which is only then absorbed by a bandgap matching photovoltaic (PV) cell. Since only the energetic portion of thermal radiation is harvested, relatively high absorber temperatures are necessary in order to yield high conversion efficiencies. Conversely, the TEPL emission, which is characterized by a blue shift in the PL spectrum, is more efficient in generating energetic photons than a thermal emitter at the same temperature [5]. In this paper, we propose and thermodynamically analyse a novel TEPL device, for the conversion of solar energy beyond the Shockley-Queisser (SQ) limit[1] at moderate operating temperatures. In addition, we present a preliminary proof-of-concept experiments, carried out with a GaAs PV cell and an $Nd^{3+}$ doped absorber, where sub bandgap PL-photons are thermally blue-shifted and harvested by the PV cell.

The thermodynamic justification for replacing a thermal absorber/emitter with a PL material is entropic: A thermal absorber must satisfy only energy conservation, and generate maximal entropy. In contrast, a PL absorber must also, by the detailed-balance principle, satisfy a rate conservation law, i.e. every absorbed photon generates a single emitted photon. This conservation in the number of photons is associated with reduced entropy generation as manifested by the existence of a non-zero chemical potential, μ[7]. The emitted entropy current can be calculated by the Gibbs free energy equation:

$$S = \frac{E-\mu N}{T} \quad (1)$$

Figure 1a shows the comparison in entropy between a PL material with bandgap, $E_g$=1 eV, and an equivalent thermal body (sketched in figure 1B), which has a step-function like emissivity. Both bodies absorb and emit the same total energy, $E$. While for a given total energy the thermal body have unique temperature and entropy, the solution for PL absorber at the total energy is a function of both μ and $T$. This can be seen in Fig. 1a. A given total energy for the thermal body sets a point at the entropy/temperature space (marked by a black circle). For the same total energy, the PL absorber has various solutions (blue equal-energy line) depending on the varying μ with T (see inset). The spectrums of solutions are all with lower entropy in comparison to the thermal body.

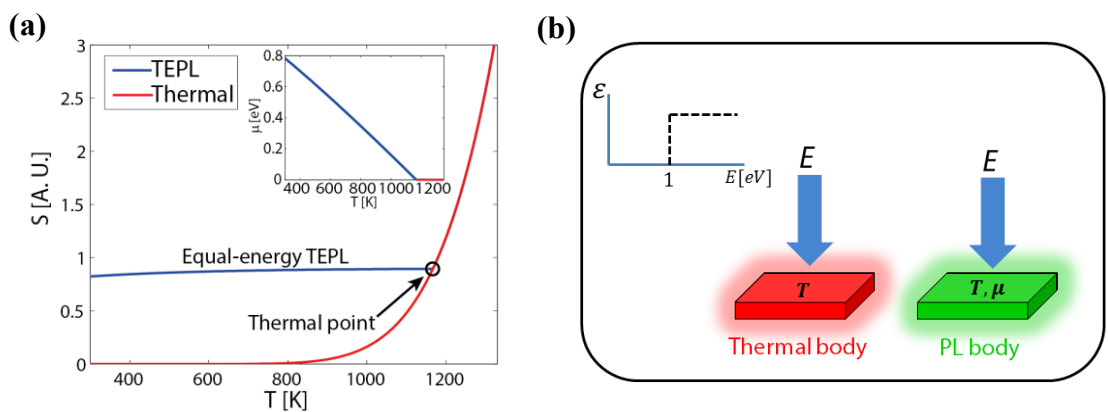

**Figure 1. (a) Entropy vs. temperature for TEPL (blue line) and thermal emissions (red line). (b) A sketch of the compared emitting bodies, with step emissivity function at the inset.**

The emitted entropy current also strongly depends on the ability to restrict emission to a specific angular content. With increasing the angular restriction under equal emitted energy, both thermal and PL emissions increase in brightness in order to satisfy the total energy balance (in the thermal and PL cases) and the balance of photon rate (only in the PL case).

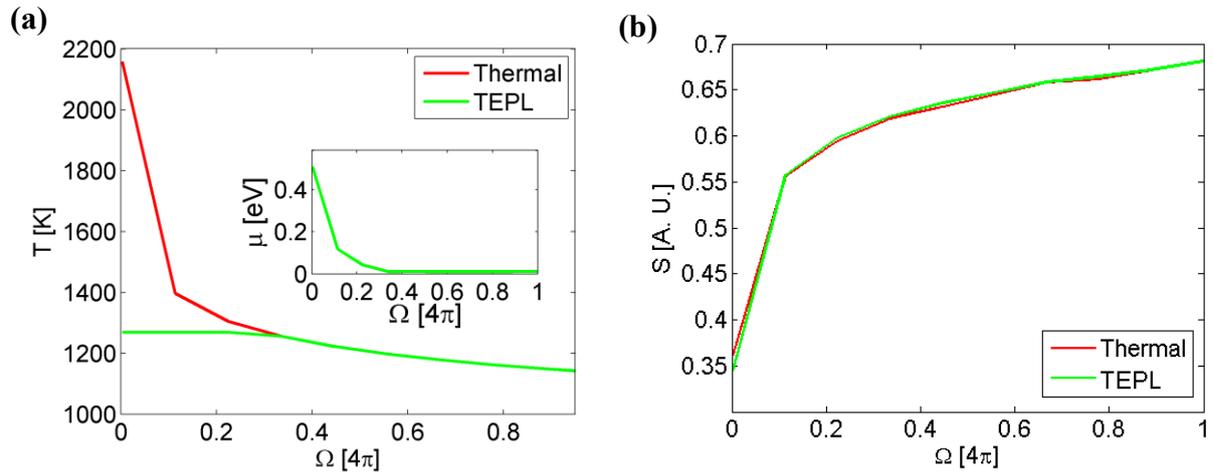

**Figure 2. (a) Thermal and TEPL body temperature dependence on solid angle of emittance. (b) Emitted entropy current dependence on solid angle**

The increase in brightness suggests a parallel decrease in entropy. Figure 2a shows the temperature dependence on the emission solid angle: when the solid angle is restricted ($\Omega$ decreases from $4\pi$ to 0), the temperature rises sharply for the thermal body, and, as long as $\mu=0$ (inset), the PL body temperature is identical. When the emission solid angle is reduced to approximately 0.35 (normalized values) the PL body's temperature stays fixed while the chemical potential rises (inset). Figure 2b shows the corresponding decrease in the emitted entropy, which is similar in both cases. Evidently, the origin of entropy decrease for the PL body is, above a critical point, the increase in $\mu$ rather than the increase in temperature, which is the origin in thermal body case.

The reduced entropy of TEPL at a fixed, lower temperature than thermal emission implies enhanced conversion efficiencies of radiation to work. The conversion efficiency analysis can also be derived from the Gibbs free energy equation:

$$\eta = \frac{E - T_L S}{E} = \frac{E - T_L \frac{E - \mu N}{T_H}}{E} = \left(1 - \frac{T_L}{T_H}\right) + \mu N \frac{T_L}{T_H} \quad (2)$$

Where $T_H$ and $T_L$ are the hot and cold temperatures and µ is the hot system's chemical potential. This term is a generalized Carnot efficiency, which reduces to Carnot's efficiency at $\mu_h=0$, i.e. the thermal body case. Figure 3 shows a comparison between the two body's conversion efficiency, with respect to a 300K cold body. It can be seen that in the regime where µ>0 (T>1300 as shown at the inset of Fig. 2a) TEPL working temperature is significantly lower than thermal body working at same efficiency. This is the main contribution of TEPL: The ability to reach similar conversion efficiencies as thermal converter while operating at significantly reduced temperatures.

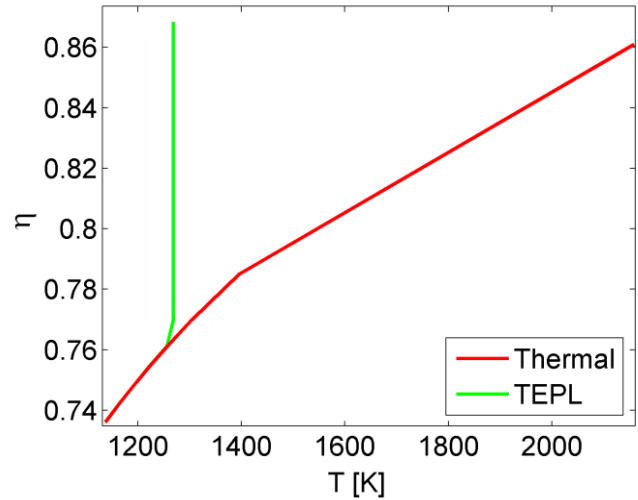

Figure 3. Carnot's efficiency limit for Thermal and TEPL emitters.

In the following section, we present and numerically analyse a TEPL device for solar energy conversion. Figure 4a shows the conceptual design. A thermally-insulated low-bandgap photoluminescent absorber completely absorbs the solar spectrum above its bandgap and emits TEPL towards a higher bandgap PV cell, maintained at room temperature. The PV converts the excessive thermal energy to electricity. For minimizing radiation losses, a semi-ellipsoidal reflective dome recycles photons by reflecting back emission at angles larger than the solid angle $\Omega_1$[8,9]. In addition, the PV's back-reflector[10,11] reflects sub-bandgap photons back to the absorber. Figure 4b shows the absorber and PV energy levels, the endothermic PL and the current flow, as well as the reflected photons from the PV. As before, the absorber's thermodynamics is governed by two conservation rules: **i.** conservation of energy. **ii.** balancing of photon flux , which takes into account the absorbed solar photons and spontaneous emission of both the absorber and the PV. At the PV, only balancing of photon flux is applied between the incoming PL, the PV's emitted light and the extracted electric current, $R_{eh}$. Energy conservation does not apply, because dissipated heat ($E_{dissipated}$) is removed from the PV to keep it at room temperature. The different fluxes are depicted in figure 4c.

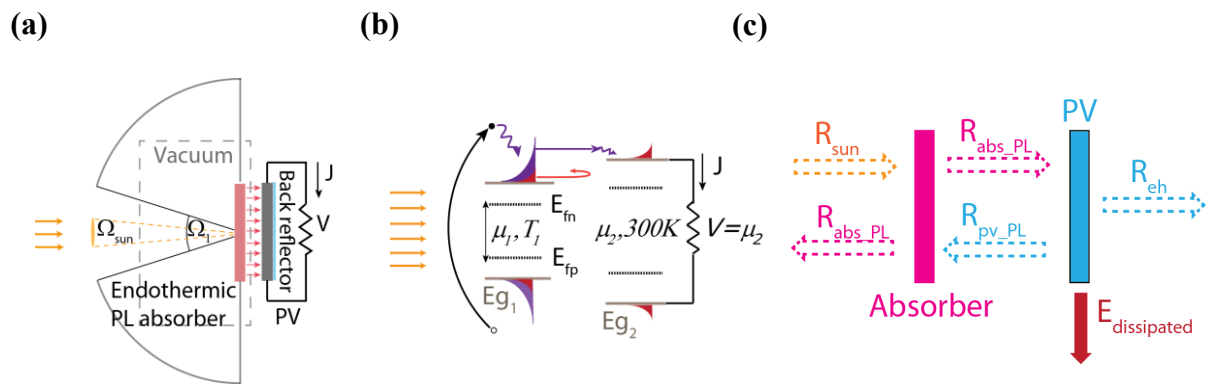

**Figure 4. (a) The device's scheme (b) The absorber and PV energy levels with the currents flow. (c) Rates included in the detailed balance (dotted arrows) and heat removal (solid red arrow)**

Assuming unity QE for both the emitter and the PV, the detailed-balance of photon and energy fluxes is uniquely accommodated by fixing the PV voltage, which is the only free parameter in the system. This defines the absorber's thermodynamic properties, $T$ and $\mu$, as well as the device's IV curve. A rigorous analysis of the detailed balance is given in the supplementary information. Theoretical maximal efficiency requires photon recycling at all angles except the acceptance solid angle of the sun ($\Omega_l=\Omega_{sun}=6.94\cdot10^{-5}$ *Srad*). However, to be more realistic we analyse the case of non-ideal photon recycling ($\Omega_l=10\Omega_{sun}$). Figure 5a depicts three different I-V curves calculated for $Eg_1=0.7\ eV$, while Eg$_2$ varies between *0.7 eV,*

*1.1 eV*, and *1.5 eV*. The I-V curve for $Eg_1= Eg_2=0.7\ eV$ (red line) shows a remarkable, "double humped" feature, and includes a thermal and a PL Maximal Power Points (MPP). While at the thermal-MPP, enhanced current leads to 53.6% efficiency, the PL-MPP remains at the SQ efficiency limit of 33% (under equivalent photon recycling). When $Eg_2$ is increased, the current enhancement vanishes and is replaced by voltage enhancement, which for $Eg_2= 1.5eV$ sets an efficiency of 65.6% (blue curve), at working temperature of 1300K. The transition between the thermal and the PL MPP's is characterized by a sharp decrease in temperature and increase in the chemical potential as elaborated at the supplementary information.

Figures 5b and 5c show the efficiency and temperature maps at the PL-MPP for various bandgap configurations assuming ideal photon recycling. For $Eg_1=Eg_2$, the efficiency is identical to the maximal SQ limit, and maximal efficiency of 70% is reached for $Eg_1=0.5eV$ and $Eg_2=1.4eV$ at T=1180K. Although the theoretical efficiencies are similar to that of STPV, operating at temperatures of 2000K-2500K[12,13], the TEPL operating temperatures are within the range of 1000K-1400K. Matching the expected maximal efficiency for the experimental parameters where the absorber is $Nd^{+3}$ with energy-gap at 1.3eV coupled to GaAs PV with bandgap at 1.45eV (cut off wavelength at 850nm) results in 45% at temperature of 740K. This assumes optimal sensitization of the $Nd^{+3}$ in order to absorb all the solar radiation at λ<900nm.

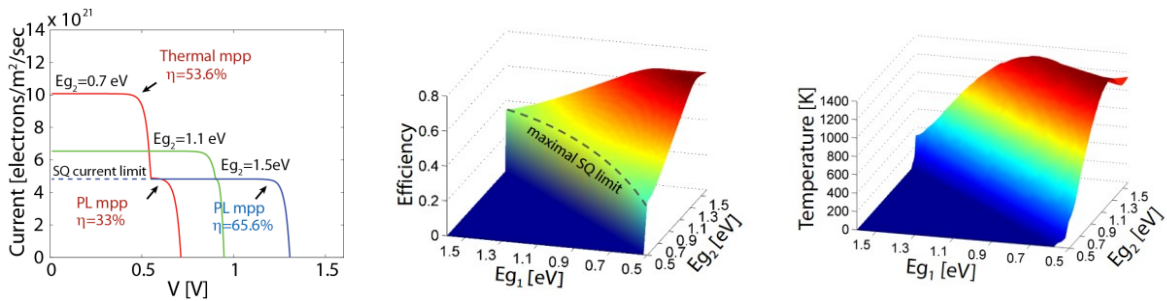

**Figure 5. Theoretical device analysis (a)** I-V curves corresponding to a *0.7 eV* absorber and PV bandgap values of (*0.7 eV, 1.1 eV* and *1.5 eV*) **(b)** Efficiency and **(c)** temperature maps, at various bandgap configurations.

From engineering considerations, it is important to arrest sub-bandgap radiative losses through phononic excitation. Such Infrared radiation can be minimized by choosing materials with low emissivity within the relevant spectral range, depending on the operating temperature. For operating at temperatures between 1200K and 1500K the peak radiative losses for emissive material are at 2.4μm-1.9μm respectively. By Kirchhoff's law of thermal radiation low emissivity materials are transparent. Therefore, matrix materials such as Silica, that are transparent at these wavelengths, are preferable choice.

We continue with an experimental demonstration of the TEPL device concept. The chosen PL system is $Nd^{3+}:SiO_2$, which was previously analysed[5]. The PL sample is mounted in a vacuum chamber, for minimization of heat losses. In order to mimic solar radiation, the sample is pumped by a 532nm laser at fluxes as high as 400 suns (1 sun= 1000w/m²). This way we are able to pump high power through the otherwise too narrow absorption bands of the $Nd^{3+}$ions. The $Nd^{3+}$ luminescence at 905nm is monitored by a calibrated spectrometer. Figure 6a shows sample's temperature increase, measured by Florescence Intensity Ratio (FIR) thermometry[5,14], as a function of the absorbed laser power calculated in suns. Figure 6b shows the corresponding evolution of the $Nd^{3+}$emission, indicating the thermally induced blue-shift, from the 900nm band to the 820nm band. This shift crosses the bandgap of GaAs at about 850nm, making these photons accessible for harvesting.

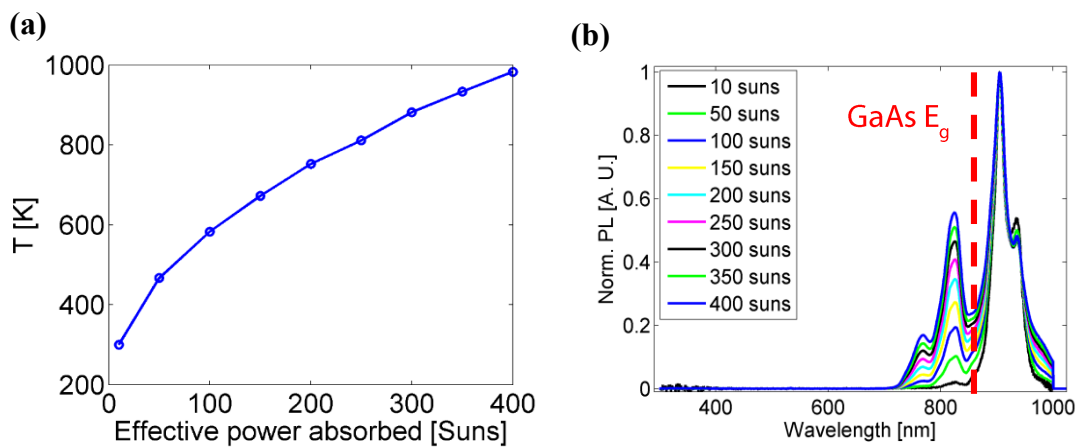

Figure 6. (a) The sample's temperature dependence on absorbed power. (b) Emission spectrum evolution

In continuation to the spectral measurements, the sample's PL was focused onto a 1 mm² GaAs cell. The cell's current was measured with a Keithley 2401 sourcemeter (Figure 7a). Figure 7b shows the photocurrent results. With the rise in absorbed power, the sample's temperature rises and the spectral shift is converted to enhanced photocurrent. As the pump power increases so does the temperature and the photocurrent. In order to demonstrate the ability of TEPL to convert photons that are sub-bandgap to the PV (unlike the 532nm photons which are initially above both the absorber and PV gaps), we pump the sample with 914nm light in parallel to the 532nm pump. The 914nm pump cannot be absorbed by the PV cell without the thermally induced blue shift. Figure 7c shows the photocurrent evolution with time, for only 532nm pump (green line), and for parallel pumps (red line). For the first case, the time evolution is similar to the one presented in figure 7b. For the parallel pump, we first turn on only the 914nm pump (indicated with an arrow), the current rises due to room-temperature blue-shift and set at a fixed level because there is no net heating of such endothermic-PL. Next we also turn on the 532nm pump. The current rises furthermore and keeps rising with time due to heating. Next the 532 pump is turned off, while the sample is still hot at ~620K, at Time~35 sec. At this point the only operating pump is the 914nm. As can be seen the photocurrent level is more than double the room temperature level.

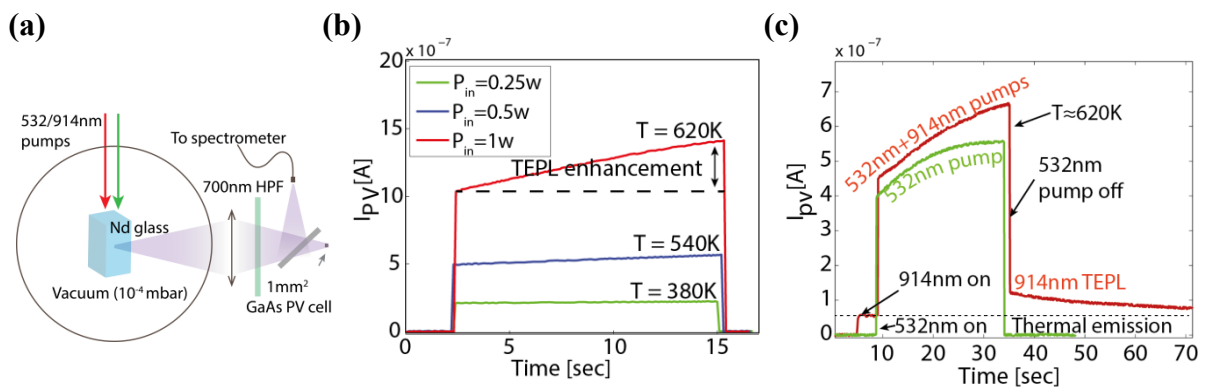

Figure 7. (a) photocurrent time-dependence at three power levels. (b) time-dependence for the parallel 914nm and 532nm pumping

This additional photocurrent is driven solely by the thermally induced blue-shift of sub-band photons. Additionally, thermal emission at similar temperature, of 620K, shows photocurrent at the noise level; orders of magnitude lower than the photocurrent generated by the TEPL conversion of 914nm pump.

To conclude, in this paper we suggested, thermodynamically analysed and demonstrated the concept of TEPL based solar energy converter. A generalized Carnot-like term for the conversion efficiency of TEPL was derived, showing that such a device is capable of performing with better conversion efficiencies, at lower operating temperatures, than a thermal emission based device. A full numerical simulation shows that an optimized TEPL device can reach efficiencies of up to 70%, in operating temperatures that are significantly lowered comparing to an STPV device. Lastly, we experimentally demonstrated the TEPL's ability to enhance PV photo-current by thermally shifting sub-bandgap radiation. This opens the way for a new direction in photovoltaics.